\documentclass[letterpaper]{article} 
\usepackage{aaai25}
\usepackage{times}  
\usepackage{helvet}  
\usepackage{courier}  
\usepackage[hyphens]{url}  
\usepackage{graphicx} 
\usepackage{natbib}  
\usepackage{caption} 
\usepackage{lipsum}
\usepackage{adjustbox}
\usepackage{xcolor}

\usepackage{algorithm}
\usepackage{algorithmic}

\usepackage{newfloat}
\usepackage{listings}
\DeclareCaptionStyle{ruled}{labelfont=normalfont,labelsep=colon,strut=off} 
\floatstyle{ruled}
\newfloat{listing}{tb}{lst}{}
\floatname{listing}{Listing}
\title{BotArtist: Generic approach for bot detection in Twitter via semi-automatic machine learning pipeline}
\author{
    Alexander Shevtsov\textsuperscript{\rm 1,2,3}, 
    Despoina Antonakaki\textsuperscript{\rm 1,2}, 
    Ioannis Lamprou\textsuperscript{\rm 1}, 
    Polyvios Pratikakis\textsuperscript{\rm 3}, 
    Sotiris Ioannidis\textsuperscript{\rm 1,2}
}
\affiliations{
    \textsuperscript{\rm 1}Technical University of Crete\\
    \textsuperscript{\rm 2}Foundation for Research and Technology Hellas - FORTH\\
    \textsuperscript{\rm 3}University of Crete, Computer Science Department\\
    shevtsov@csd.uoc.gr, dantonakaki@tuc.gr, ilamprou1@tuc.gr

}

\usepackage{bibentry}

\begin{document}

\maketitle

\begin{abstract}
Twitter, as one of the most popular social networks, provides a platform for communication and online discourse. Unfortunately, it has also become a target for bots and fake accounts, resulting in the spread of false information and manipulation. This paper introduces a semi-automatic machine learning pipeline (SAMLP) designed to address the challenges associated with machine learning model development. Through this pipeline, we develop a comprehensive bot detection model named BotArtist, based on user profile features. \\
SAMLP leverages nine distinct publicly available datasets to train the BotArtist model. To assess BotArtist's performance against current state-of-the-art solutions, we evaluate 35 existing Twitter bot detection methods, each utilizing a diverse range of features. Our comparative evaluation of BotArtist and these existing methods, conducted across nine public datasets under standardized conditions, reveals that the proposed model outperforms existing solutions by almost 10\% in terms of F1-score, achieving an average score of 83.19\% and 68.5\% over specific and general approaches, respectively.\\
As a result of this research, we provide one of the largest labeled Twitter bot datasets. The dataset contains extracted features combined with BotArtist predictions for 10,929,533 Twitter user profiles, collected via Twitter API during the 2022 Russo-Ukrainian War over a 16-month period. This dataset was created based on \citep{shevtsov2022twitter} where the original authors share anonymized tweets discussing the Russo-Ukrainian war, totaling 127,275,386 tweets. The combination of the existing textual dataset and the provided labeled bot and human profiles will enable future development of more advanced bot detection large language models in the post-Twitter API era.
\end{abstract}

%

\section{Introduction}

Online social media has become an essential part of everyday life. During the past decade, social networks have transformed the communication routine of our daily lives. Due to their growing popularity, online social media gained millions of daily active users not only consuming the information but also creating a space for content creators. The main reason behind the success of online social networks is the real-time access to unlimited information, where registered users can share their comments and personal opinions on popular topics.

Twitter, one of the most popular social networks, with millions of active users, is used for news dissemination, political discussions, and social interactions. However, the platform has also been plagued by bots and fake accounts that are used to manipulate and spread false information. According to the research community, the usage of manipulation techniques implemented with the use of bot accounts is registered during diverse popular topic discussions. More specifically, studies show that bot accounts are involved in diverse political discussions, for instance, around 2016 in countries like the US, Germany, Sweden, France, Spain, etc. \citep{golovchenko2020cross,badawy2018analyzing,howard2016bots,shevtsov2022identification,shevtsov2023tweets,neudert2017junk,pastor2020spotting,bradshaw2017junk,fernquist2018political,castillo2019detection,rossi2020detecting}. Furthermore, propagandizing and false news is identified during the vaccination debate \citep{broniatowski2018weaponized} and more recent examples of the COVID-19 pandemic \citep{shahi2021exploratory, ferrara2020types, yang2021covid}. This high activity of bot accounts raises concerns in the research community and online social media platforms about the integrity of the shared information.

Currently, the research community offers a variety of ML and NN-based bot detection methods. Each of these provided methods excels at addressing specific bot detection scenarios, yielding optimal solutions for particular use cases. Unfortunately, the performance of these provided methods significantly diminishes in more general bot detection scenarios, encompassing different periods, discussion topics, and languages, among others. Recent studies have shown that existing methods still do not achieve flawless Twitter bot detection, regardless of the number of user characteristics that they incorporate \citep{feng2022twibot}. In addition, a large portion of existing ML bot detection methods ignore well-known optimization procedures (including feature selection and dynamic hyper-parameter fine-tuning) providing significant space for improvement.

In our research, we address this challenge by developing a semi-automatic machine learning pipeline (SAMLP\footnote{GitHub repository: 
https://github.com/alexdrk14/SAMLP
}) for constructing a generic bot detection model (BotArtist\footnote{GitHub repository: 
https://github.com/alexdrk14/BotArtist}). The presented pipeline resolves popular issues during the creation of the classification model, including recursive hyperparameter fine-tuning during the feature selection phase. This approach guarantees the reduction of non-relevant features which sequentially reduces the noise of initial data. Additionally, our pipeline considers the class imbalance during feature selection and model fine-tuning, making the model suitable for real-case applications.
Next, we compare this developed model with 35 currently available state-of-the-art approaches using nine publicly available datasets. This research yields two distinct evaluation scenarios: one specific to each dataset, where each model is trained and tested separately on each of them, and a more generic approach, where the models are trained and tested on the combined dataset comprising all nine unique datasets.

As a result of our research, we introduce BotArtist, a finely tuned general bot detection model that achieves the highest F1-score performance on three of the nine datasets, with an average F1-score performance of 83.19. Furthermore, the presented approach achieves greater precision in general evaluation, with an overall improvement in F1 score of more than 9\% and an average improvement of 5\% compared to the best-performing existing bot detection approaches. Moreover, BotArtist relies on only a limited set of profile features, allowing for the prediction of historical data independently of the Twitter API.

Except from the source code of SAMLP and BotArtist, we make available the model outcome usable for research. For this purpose, we collect a large volume of users correlated to the already existing textual shared dataset and add an additional layer to textual information via user prediction. We then release 10.929.533 user profiles with BotArtist predictions \footnote{Zenodo repository: https://zenodo.org/records/11203900}, correlated to 127.275.386 tweets \citep{shevtsov2022twitter}.

\begin{table*}[tb]
    \centering
    
    \begin{tabular}{c|c|c|c|c|c|c|c|c|c}
    \hline
         {\bfseries Dataset} &  {\bfseries C-15} & {\bfseries G-17} & {\bfseries C-17} & {\bfseries M-18} & {\bfseries C-S-18} & {\bfseries C-R-19} & {\bfseries B-F-19} & {\bfseries TwiBot-20} & {\bfseries TwiBot-22} \\ \hline \hline
         {\bfseries \# Total User} & 5,301 & 2,484 & 14,368 & 50,538 & 13,276 & 693 & 518 & 229,580 & 1,000,000 \\
         {\bfseries \# Human} & 1,950 & 1,394 & 3,474 & 8,092 & 6,174 & 340 & 380 & 5,237 & 860,057 \\ 
         {\bfseries \# Bot} & 3,351 & 1,090 & 10,894 & 42,446 & 7,102 & 353 & 138 & 6,589 & 139,943 \\ \hline
         
         {\bfseries \# Total Tweet} & 2,827,757 & 0 & 6,637,615 & 0 & 0 & 0 & 0 & 33,488,192 & 88,217,457 \\
         {\bfseries \# Human Tweet} & 2,631,730 & 0 & 2,839,361 & 0 & 0 & 0 & 0 & 927,292 & 81,250,102 \\
         {\bfseries \# Bot Tweet} & 196,027 & 0 & 3,798,254 & 0 & 0 & 0 & 0 & 1,072,496 & 6,967,355 \\ \hline
         {\bfseries \# Graph Edges} & 7,086,134 & 0 & 6,637,615 & 0 & 0 & 0 & 0 & 33,488,192 & 170,185,937 \\ \hline 
    \end{tabular}
    \caption{Description of selected datasets and information contained in those datasets.}
    \label{tab:datasets}
    
\end{table*}

\section{Related Work}

Bot detection on Twitter poses a formidable challenging task, primarily due to the increasing sophistication of these bots. Research efforts in this domain can generally be categorized into three main approaches: feature-based, text-based, and graph-based techniques. Each category offers a distinct angle on bot detection, leveraging various extracted characteristics of users and their activities on the social media platform.

\subsection{Feature based}

Methods falling into this category apply feature engineering based on information extracted from Twitter user profiles and their activity patterns. Researchers use traditional machine learning or neural network classification algorithms for bot detection. These approaches employ different sets of user characteristics as feature sets, including user metadata \citep{kudugunta2018deep}, tweets \citep{miller2014twitter}, user name \citep{beskow2019its}, description \citep{hayawi2022deeprobot}, temporal patterns \citep{mazza2019rtbust}, and follow relationships \citep{feng2021satar}. Moreover, some of them aim to enhance the scalability of feature-based methods \citep{yang2020scalable,abreu2020twitter}, discover unknown bot accounts using correlation-based techniques \citep{chavoshi2016debot}, and improve the trade-off between precision and recall in bot detection \citep{morstatter2016new}. However, the creators of bot accounts are increasingly aware of the features utilized by the research community, leading to novel bot implementations designed to evade detection based on known features \citep{cresci2020decade}. Consequently, existing feature-based methods face several challenges in the accurate detection of these new bot accounts \citep{feng2021satar}.

\subsection{Text based}

Text-based methods rely on natural language processing (NLP) techniques for Twitter bot detection, extracting characteristics from posted content (tweets) and user profile descriptions. Approaches in this category include sequence fingerprint \citep{cresci2016dna}, word embeddings \citep{wei2019twitter}, recurrent neural networks (RNN) \citep{kudugunta2018deep}, attention mechanisms \citep{feng2021satar}, transformers \citep{guo2021social,liu2019roberta,raffel2020exploring} and the adoption of pre-trained language models for encoding tweets \citep{dukic2020you}. Researchers have also combined tweet representations with user profile characteristics \citep{cai2017behavior,efthimion2018supervised,kantepe2017preprocessing,miller2014twitter,varol2017online,kouvela2020bot,lee2011seven,echeverri2018lobo}, employed unsupervised machine learning models \citep{feng2021satar}, and addressed multilingual context issues \citep{knauth2019language}. Despite extensive existing research studies and impressive performance in text-based approaches, new bot accounts can still evade detection by sharing stolen content from genuine users \citep{cresci2020decade}. Additionally, recent work has demonstrated that relying solely on textual information is insufficient for robust and accurate bot detection \citep{feng2021botrgcn}.

\subsection{Graph based}
The graph-based category of bot detection methods combines geometric deep neural networks with graph analytics. Current implementations leverage techniques such as node centrality \citep{dehghan2023detecting}, node representation learning \citep{pham2022bot2vec}, graph neural networks (GNNs) \citep{ali2019detect,moghaddam2022friendship}, and heterogeneous GNNs \citep{feng2021botrgcn} to conduct graph-based bot detection. In recent research, the authors have borrowed ideas from other categories to merge different approaches, such as combining multiple methods \citep{guo2021dual,yang2022botometer,knauth2019language,beskow2020you,rodriguez2020one,magelinski2020graph,yang2013empirical,kipf2016semi,kipf2016semi,velivckovic2017graph,lv2021we}. Furthermore, novel GNN architectures have been proposed to exploit heterogeneity in the Twitter network \citep{feng2022heterogeneity}. Generally, these approaches hold significant promise for Twitter bot detection.

Unfortunately, existing Twitter bot detection methods have their limitations. Simplified feature-based approaches struggle to generalize and provide robust results, while more complex methods based on text or graph characteristics require intensive computational resources and large datasets for model development. Furthermore, as shown in \citep{shevtsov2023russo} the complex embedding-based model tends to perfectly capture the training data patterns, but lacks generalizability and requires frequent retraining over multiple datasets to keep high performance.
These limitations, coupled with the recent announcement of the Twitter API monetization \citep{twitterAPI2023}, make many existing methods expensive to maintain or nearly impossible to operate daily. In this study, we address these challenges and present robust and accurate Twitter bot detection methods, based on a lightweight set of features. Our methods require only a single user profile object (Twitter API v1.1 or v2) to provide user prediction without additional Twitter API requests, thereby significantly reducing operational costs.

\begin{figure*}[tb]
    \centering
    \includegraphics[width=0.8\textwidth]{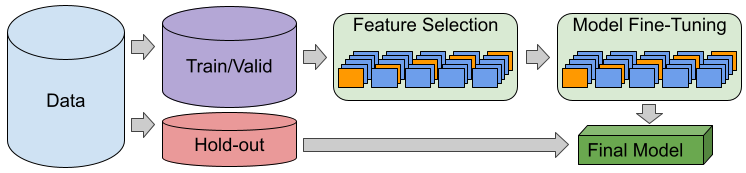}
    \caption{Machine learning pipeline used for \textit{BotArtist} model creation including feature selection and model fine-tuning; each step is executed during separate K-Fold cross-validation.}
    \label{fig:pipeline}
\end{figure*}
 
\section{Datasets}

For this research paper, we collect nine well-known publicly available datasets \citep{feng2021twibot}. All selected datasets already contain ground truth labels, primarily obtained through manual analysis or crowd-sourcing. For simplicity, we label the selected datasets as follows: C-15 \citep{cresci2015fame}, G-17 \citep{gilani2017bots}, C-17 \citep{cresci2017social,cresci2017paradigm}, M-18 \citep{yang2020scalable}, C-S-18 \citep{cresci2018fake,cresci2019cashtag}, C-R-19 \citep{mazza2019rtbust}, B-F-19 \citep{yang2019arming}, TwiBot-20 \citep{feng2021twibot}, and TwiBot-22 \citep{feng2022twibot}. In Table \ref{tab:datasets}, we present the information provided in each dataset, along with the volume of normal and bot accounts.

Furthermore, in collaboration with \citep{shevtsov2022twitter} collect 10.929.533 Twitter profiles correlated with 127.275.386 publicly available tweets related to the public discussion topic of the 2022 Russo-Ukrainian War. Our collection of user profiles is based on the monitoring of selected topics starting from February 23, 2022, till June 23, 2023. The shared datasets contain a set of extracted features in an anonymized form of a CSV file and contain only preprocessed numerical features to protect user information. The provided dataset also provides anonymized user IDs which are identically correlated with publicly available user tweet dataset \citep{shevtsov2022twitter}.


\section{Methodology}

The proposed approach is based on the development of a semi-automatic machine learning pipeline (SAMLP). We choose this implementation due to its simplicity for further usage and its ability to avoid many trivial mistakes during the ML model development such as data processing, feature selection, hyper-parameter fine-tuning, and model evaluation. The steps of the developed pipeline are presented in Figure \ref{fig:pipeline}, where the initial step involves data splitting.

During data splitting, it is crucial to maintain a class balance between the training/validation and testing data portions. 
For this purpose, we apply a stratified data split with a 70:30 ratio for training, validation, and testing portions respectively. Such an approach allows us to preserve the natural class distribution difference.
The testing data portion is kept hidden and is only utilized during the final testing of the model to prevent information leakage between train and test portions

\subsection{Feature Selection}
After the data splitting, our pipeline applies a feature selection procedure over the train/validation data portion during the K-Fold cross validation. 
Although K-Fold cross-validation may lead to slightly over-optimistic performance estimations, in our case, it is not relevant since we use this procedure for $\alpha$ hyper-parameter fine-tuning of the Lasso model, and we are interested in identifying the best $\alpha$ parameter.

Additionally, since the Lasso regression model does not perform well over imbalanced datasets, we also take into account the class imbalance of the 
dataset. In case of data imbalance,
we utilize under-sampling of the majority class, making the data perfectly balanced and improving the prediction performance of the Lasso model. Due to the high-class imbalance and the under-sampling of the majority class, there is a very high probability of losing some important samples, which may lead to inaccurate feature selection. To address this, we utilize 10 consecutive repetitions to capture as many samples of the majority class as possible.

During these repetition rounds, we store the best $\alpha$ parameters based on the square mean error metric from the entire stratified k-Fold cross-validation results, where $K=5$. At the end of the procedure, we compute the most frequent best $\alpha$ parameter and train the Lasso model with the entire train/validation dataset, keeping features with non-zero coefficients. Additionally, we check if the selected $\alpha$ parameter falls within the defined limits of the searching area. If it falls on the minimum or maximum values, we create a new search area based on the original site to ensure the selected $\alpha$ is genuinely the best.

This approach allows us to utilize feature selection without additional knowledge of the original data and modifications, as the procedure automatically finds the best $\alpha$ values, manages class balance and imbalance via under-sampling and repeated executions, and processes both perfectly balanced and highly imbalanced datasets without information loss.

\subsection{Model Fine-Tuning}
Having reduced the problem's dimensionality through feature selection, we now focus on identifying the model and its configuration (hyper-parameter set) that can provide accurate predictions over unseen data samples. We select three well-known machine learning classification models: SVM \citep{boser1992training}, RandomForest \citep{breiman2001random}, and state-of-the-art XGBoost (Extreme Gradient Boosting) \citep{chen2015xgboost}.

For each model, we define large hyper-parameter ranges to cover a variety of configurations. To reduce complexity and execution time, we develop a sampling method that randomly selects $C$ unique configurations per model (in our case, $C=50$). This method allows us to estimate different possible configurations without sacrificing execution time. Additionally, since the dataset may be affected by class imbalance, we compute class-specific weights to construct selected classifiers with consideration of class imbalance.

To evaluate each selected configuration, we use a stratified K-Fold cross-validation approach, allowing us to utilize the entire train/validation data portion and evaluate each configuration over multiple validation data portions. During the evaluation, we compute the average configuration performance measured in the F1-score, which can effectively measure model performance over both binary and multi-class classification tasks, considering class imbalance.

Afterward, we select the best classifier and configuration based on the average validation (F1-score). This model configuration becomes the final model, which will be evaluated over the hold-out (testing) data portion. In the final stage of the model development, we provide model explainability using the SHAP game-theoretical approach  \citep{Shapley1953}. This allows us to describe the reasons behind the model's decisions and highlight the distinctions between classes.

Additionally, in the case of binary classification, we need to adjust the decision threshold, as binary classification models are not perfectly aligned with the 50\% decision threshold. To do this, we utilize the precision vs. recall curve to identify the decision threshold that maximizes the model's performance. The developed model is then trained over the entire dataset, including the train/validation and testing sets, and can be utilized for real-world applications with the identified decision threshold.

Described experiments were conducted in the single machine with AMD Ryzen 9 CPU (16 cores/32 threads) with a total of 64 GB DDR4 system memory. For the graphical unit based computations the single unit of Nvidia RTX 3080 where used with total memory of 12 GB.

\begin{table*}[tb!]
    \centering
    
    \begin{tabular}{|c c||c c c|}
    \hline
        \textbf{Feature} & \textbf{Type} & \textbf{Feature} & \textbf{Type} & \textbf{Calculation}\\ \hline \hline
        name\_len & count & screen\_name\_sim & real-valued & Jaccard of (screen\_name, user\_name)   \\ \hline
        screen\_name\_len & count & foll\_friends  & real-valued & follower / friends \\ \hline 
        description\_len & count & age & real-valued & created\_at / collection date\\ \hline

        listed & count & listed\_by\_age & real-valued & listed / account age\\ \hline
        statuses & count & statuses\_by\_age & real-valued & statuses / account age\\ \hline
        followers & count & followers\_by\_age & real-valued & followers / account age\\ \hline
        following & count & following\_by\_age & real-valued & friends / account age\\ \hline 

        name\_upper\_len & count & name\_upper\_pcnt & real-valued & percentage of upper case\\ \hline
        name\_lower\_len & count & name\_lower\_pcnt & real-valued & percentage of lower case\\ \hline
        name\_digits\_len & count & name\_digits\_pcnt & real-valued & percentage of digits\\ \hline
        name\_special\_len & count & name\_special\_pcnt & real-valued & percentage of other characters\\ \hline
        screen\_name\_upper\_len & count & screen\_name\_upper\_pcnt & real-valued & percentage of upper case\\ \hline
        screen\_name\_lower\_len & count & screen\_name\_lower\_pcnt & real-valued & percentage of lower case\\ \hline
        screen\_name\_digits\_len & count & screen\_name\_digits\_pcnt & real-valued & percentage of digits\\ \hline
        screen\_name\_special\_len & count & screen\_name\_special\_pcnt & real-valued & percentage of other characters\\ \hline
        description\_upper\_len & count & description\_upper\_pcnt & real-valued & percentage of upper case\\ \hline
        description\_lower\_len & count & description\_lower\_pcnt & real-valued & percentage of lower case\\ \hline
        description\_digits\_len & count & description\_digits\_pcnt & real-valued & percentage of digits\\ \hline
        description\_special\_len & count & description\_special\_pcnt & real-valued & percentage of other characters\\ \hline
        description\_urls & count &  name\_entropy & real-valued & entropy of user name\\ \hline
        description\_mentions & count & screen\_name\_entropy & real-valued & entropy of the screen name\\ \hline
        description\_hashtags & count &  has\_location	 & boolean & \\ \hline
        
        total\_urls & count & has\_profile\_image & boolean  & \\ \hline
        protected & boolean &  has\_profile\_url & boolean & \\ \hline
        verified & boolean &  & &\\ \hline

    \end{tabular}
    \caption{List of the profile features extracted from the Twitter API user objects.}
    \label{tab:features}
    
\end{table*}

\subsection{Feature Extraction}

As mentioned earlier, in this study we pursue the creation of a simple model that in comparison with other bot detection systems- does not require a large volume of information or features for accurate prediction. In our case, the number of Twitter profiles triggers our attention. The majority of social media bot accounts are created for content promotion and increasing the attention of other registered users. To maximize this goal, bot accounts aim for the expansion of social audience. Social expansion is very difficult to achieve without high activity, such as tweets, re-tweets, comments, etc. In comparison with profile-based features, the extraction and utilization of textual or graph features can result in instability, causing a significant performance drop across various periods or discussion topic datasets \citep{shevtsov2023russo}. Consequently, we need to extract as much information as possible from the user profiles to identify extremely active accounts. The difference between very popular social media accounts (also known as celebrities), and bot accounts is that they try to post and share as much content as possible during the day. 
Due to the differences between Twitter API v1.1 and v2, we have devised a method for constructing features in a way that allows the extraction of selected information from both v1.1 and v2 without any additional manipulation. The profile features we have extracted are categorized into three distinct groups: count, boolean, and real-valued.

The initial count feature categories encompass raw Twitter API object values, including followers, followings, status, and number of subscribed Twitter lists. In addition to these raw values, we calculate statistics for textual fields such as user name, screen\_name, and description. For these features, we quantify the number of distinct characters, including uppercase letters, lowercase letters, digits, and special characters that do not fall into any of the previous categories. Furthermore, we leverage the profile description field, where users can provide additional information such as mentions of other users/organizations, hashtags, and additional URLs. In this context, we have counted the volume of the mentioned entities in the description field as raw values.

In contrast to the count categories, where values are represented as integers, the real-valued categories contain floating-point values. We conduct various comparisons and measurements across the profile fields in this category. Initially, we compute the age of the accounts, which allows us to distinguish highly active users with millions of posts spanning several years from users with similar activity but with an active period of only a few days. Account age is measured in days, calculated as the duration between the account creation date, provided by the Twitter API, and the collection date. Based on the account's age, we can compute metrics similar to those in the count values category but adjusted for the account's age perspective. 


Given that many bot accounts are mass-created by automated scripts, they often have very similar or even trivial user names and screen names. Therefore, we apply the Jaccard similarity measure between user names and screen names to calculate the intersection of characters used in both fields to the differences between them.


Additionally, we compute the entropy of the user name and screen\_name, enabling us to assess the randomness of these selected fields. In addition to the count values derived from the profile text fields (user name, screen\_name, and description), we compute the percentage of each specific character category, including lowercase, uppercase, digits, and special characters, relative to the length of the text field. These measurements offer a more precise means of identifying textual preferences more flexible to varying text lengths since they are computed as percentages of the field length.

The final category of extracted features consists of Boolean values, used to identify information that may or may not be provided by the account creator. For instance, we determine if the user account is protected, and verified, as well as extract information about the provided user location, profile URL, and whether the profile uses the default profile image.

By employing this comprehensive feature engineering approach, we maximize the feature extraction while working with the limited profile information available through Twitter API v2. In total, we extract 49 unique profile features Table \ref{tab:features}, representing the full set of features used before the feature selection process in the BotArtist model.

\begin{table*}[htb!]
    \centering
    
    \begin{tabular}{|c|c|ccccccccc||c|}
    \hline
    \bf{Method} & \bf{Type} &\bf{C-15}    & \bf{G-17}  & \bf{C-17}   & \bf{M-18}   & \bf{C-S-18} & \bf{C-R-19} & \bf{B-F-19} & \bf{TB-20} & \bf{TB-22}  & \bf{Average}\\ \hline \hline
    SGBot           & F & 77.9            & 72.1 & 94.6            & 99.5            & 82.3 & 82.7            & 49.6            & 84.9 & 36.6 & 75.57\\ 
    Kudugunta et al.& F & 75.3            & 49.8            & 91.7            & 94.5            & 50.9            & 49.2            & 49.6            & 47.3             & 51.7 & 62.22\\ 
    Hayawi et al.   & F & 85.6            & 34.7            & 93.8            & 91.5            & 60.8            & 60.9            & 20.5            & 77.1             & 24.7 & 61.06\\ 
    BotHunter       & F & 97.2 & 69.2            & 91.6            & \underline{99.6}& 82.2& 82.9            & 49.6            & 79.1             & 23.5 & 74.98\\ 
    NameBot         & F & 83.4            & 44.8            & 85.7            & 91.6            & 61.1            & 67.5            & 38.5            & 65.1             & 0.5 & 59.80\\ 
    Abreu et al.    & F & 76.4            & 66.7            & 95.0 & 97.9            & 76.9            & \underline{83.5}& \underline{53.8}& 77.1             & 53.4 & 75.63\\ 
    BotArtist       & F & 98.3              & \underline{76.1}               & 97.0               & {\bfseries99.7} & 80.6            & {\bfseries88.3} & {\bfseries68.4} & 82.2 & \underline{58.2}  & {\bfseries83.19}\\ \hline \hline

    Cresci & T & 1.17 & - & 22.8 & - & - & - & - & 13.7 & - & -  \\ 
    Wei    & T & 82.7 & - & 78.4 & - & - & - & - & 57.3 & 53.6 & - \\ 
    BGSRD  & T & 90.8 & 35.7 & 86.3 & 90.5 & 58.2 & 41.1 & 13.0 & 70.0 & 21.1 & 56.30\\ 
    RoBERTa & T & 95.8 & - & 94.3 & - & - & - & - & 73.1 & 20.5 & - \\ 
    T5 & T & 89.3 & - & 92.3 & - & - & - & - & 70.5 & 20.2 & - \\ \hline \hline 
    
    Efthimion & FT & 94.1 & 5.2 & 91.8 & 95.9 & 68.2 & 71.7 & 0.0 & 67.2 & 27.5 & 57.95\\ 
    Kantepe & FT & 78.2 & - & 79.4 & - & - & - & - & 62.2 & {\bfseries58.7} & -\\
    Miller & FT & 83.8 & 59.9 & 86.8 & 91.1 & 56.8 & 43.6 & 0.0 & 74.8 & 45.3 & 60.23 \\
    Varol & FT & 94.7 & - & - & - & - & - & - & 81.1 & 27.5 & -\\
    Kouvela & FT & 98.2& 66.6 & \underline{99.1} & 98.2 & 80.4 & 81.1 & 28.1 & 86.5 & 30.0 & 74.24\\
    Santos & FT & 78.8 & 14.5 & 83.0 & 92.4 & 65.2 & 75.7 & 21.0 & 60.3 & - & -\\
    Lee & FT & \underline{98.6} & 67.8 & {\bfseries99.3} & 97.9 & \underline{82.5} & 82.7 & 50.3 & 80.0 & 30.4 & \underline{76.61}\\
    LOBO & FT & {\bfseries98.8}& - & 97.7 & - & - & - & - & 80.8 & 38.6 & -\\ \hline \hline

    Moghaddam & FG & 73.9 & - & - & - & - & - & - & 79.9 & 32.1 & -\\
    Alhosseini & FG & 92.2 & - & - & - & - & - & - & 72.0 & 38.1 & -\\ \hline \hline

    Knauth & FTG & 91.2 & 39.1 & 93.4 & 91.3 & {\bfseries94.0} & 54.2 & 41.3 & 85.2 & 37.1 & 69.64\\
    FriendBot & FTG & 97.6 & - & 87.4 & - & - & - & - & 80.0 & - & -\\
    SATAR & FTG & 95.0 & - & - & - & - & - & - & 86.1 & - & -\\
    Botometer & FTG & 66.9 & {\bfseries77.4} & 96.1 & 46.0 & 79.6 & 79.0 & 30.8 & 53.1 & 42.8 & 63.5\\
    Rodrifuez-Ruiz & FTG & 87.7 & - & 85.7 & - & - & - & - & 63.1 & 56.6 & -\\
    GraphHist & FTG & 84.5 & - & - & - & - & - & - & 67.6 & - & -\\
    EvolveBot & FTG & 90.1 & - & - & - & - & - & - & 69.7 & 14.1 & -\\
    Dehghan & FTG & 88.3 & - & - & - & - & - & - & 76.2 & - & -\\
    GCN & FTG & 97.2 & - & - & - & - & - & - & 80.8 & 54.9 & -\\
    GAT & FTG & 97.6 & - & - & - & - & - & - & 85.2 & 55.8 & -\\
    HGT & FTG & 96.9 & - & - & - & - & - & - & {\bfseries88.2} & 39.6 & -\\
    SimpleHGN & FTG & 97.5 & - & - & - & - & - & - & {\bfseries88.2} & 45.4 & -\\
    BotRGCN & FTG & 97.3 & - & - & - & - & - & - & 87.3 & 57.5 & -\\
    RGT & FTG & 97.8 & - & - & - & - & - & - & \underline{88.0} & 42.9 & -\\ \hline
    \end{tabular}
    \caption{The performance of each selected bot detection model, as reported in the \citep{feng2021twibot} paper, is compared with that of BotArtist. Performance is measured using the F1-score. In this benchmark, each model is trained and tested on each dataset separately.}
    \label{tab:f1_single_scores}
    
\end{table*}

\section{Experimental Results}\label{sec:experiments}

According to the methodology of SAMLP described earlier, we successfully develop a semi-automated machine learning pipeline capable of constructing classification models (both binary and multiclass) for both balanced and imbalanced datasets. Using this pipeline, we create and test our custom implementation called BotArtist for Twitter bot detection, utilizing nine well-known datasets. To evaluate BotArtist's performance, in comparison to 35 other existing research approaches, we employ the most comprehensive Twitter bot detection benchmark available \citep{feng2022twibot}, where various methods are compared using identical training/validation and testing data portions. This benchmarking allows us to make a fair comparison between our approach and established bot detection methods.

In our comparison, we not only evaluate BotArtist against bot detection approaches based on similar categories of features but also compare it to existing approaches developed based on different sets of characteristics. The selected methods are categorized into five different groups: feature-based (F), textual (T), graph (G), and any combination of them.

\begin{table*}[htb!]
    \centering
    
    \begin{tabular}{|c|c|c|c|c|c|c|c|c|c||c|c|}
    \hline
    \textbf{Method} & C-15    & G-17  & C-17 & M-18 & C-S-18 & C-R-19 & B-F-19 & TB-20 & TB-22 & Total & Average\\ \hline \hline
    BotArtist       & 82.7    & \underline{39.9}  & 87.3 & \underline{99.0}& {\bfseries80.6} & \underline{73.8} & 16.6 & {\bfseries80.3} & {\bfseries56.9} & {\bfseries63.7} & {\bfseries68.5}\\
    Lee             & 82.3    & 0.0   & 83.6 & 97.7 & 78.2   & 67.7   & 20.0 & 8.5 & 42.4 & 52.9 & 53.3\\ 
    Abreu           & \underline{84.4}    & 0.3   & 80.1 & 88.4 & 67.1   & 40.8   & 11.7 & 15.6 & 29.0 & 40.4 & 46.3\\ 
    SGBot          & 75.0    & 3.6   & 79.8 & {\bfseries99.2} & 76.7   & 68.9   & 0.0 & 15.2 & \underline{43.3} & \underline{53.8} & 51.5\\
    BotHunter      & 73.4    & 7.1   & 76.0 & {\bfseries99.2} & 76.0   & 44.8   & 11.1 & 14.7 & 28.0 & 43.1 & 47.8\\ 
    Kouvela & {\bfseries95.5} & 20.4 & \underline{94.7} & 98.1 & 78.4 & 71.4 & \underline{21.0} & 28.5 & 36.0 & 52.0 & 60.5\\
    Botometer       & 66.9    & {\bfseries77.4}  & {\bfseries96.1} & 46.0 & \underline{79.6} & {\bfseries79.0} & {\bfseries30.8} & \underline{53.1} & 42.8 & 45.3 & \underline{63.5}\\  \hline   
    \end{tabular}
    \caption{The measurement of performance in the case of general bot detection approaches, involves training and testing models on all datasets. Performance is assessed using the F1-score.}
    \label{tab:f1_all_scores}
\end{table*} 

\subsection{Models comparison}

Our goal with BotArtist is to create a model with the highest possible level of generalizability. To achieve this, we design two separate comparison scenarios. Initially, our focus is on assessing the fine-tuned BotArtist's ability to capture general patterns within each dataset in our collection. As a result, we conduct training and testing for each of the selected models, including BotArtist, on each of the nine chosen datasets individually. This comparative approach allows us to evaluate how well these selected models perform on each specific dataset, serving as a more rigorous bot detection implementation.

Additionally, the designed experiments provide us with a broader understanding of the overall capabilities of the selected methods when applied in more general scenarios. While implementing these experiments, we take into account that the performance would likely be higher than that of a generic approach. Therefore, our primary focus is on discerning the differences between the models to select the most accurate bot detection methods for the broader, more generalized comparison. Based on the designed experiment we identified that (see Table \ref{tab:f1_single_scores}) the BotArtist model emerged as the only bot detection method that achieves superior performance across three different datasets (M-18, C-R-19, B-F-19). Furthermore, the BotArtist model demonstrates the highest average F1-score, reaching 83.19. This represents a substantial improvement of 6.5\% over the most accurate of the existing methods. As shown in Table \ref{tab:f1_single_scores}, some datasets posed challenges for more complex methods that rely on text and graph features. These methods struggle to utilize datasets without text or graph information due to their limitations in accessing essential information required for the model.

Previously, we had evaluated and compared models on each isolated dataset and calculated their average performance. While this approach provided us with an initial insight into model performance across diverse datasets, it lacked an assessment of model generalizability. To address this, we design additional comparisons where we assess the model's ability to capture general patterns across different datasets. In these cases, we combine the training data from the nine datasets into a single dataset for model training. Combining data from multiple datasets, each with varying periods, discussion topics, and communities, demands a high level of model generalization to avoid overfitting. 

To measure the performance of the trained methods, we evaluate them by testing on portions of each dataset and also on the union of all the dataset testing portions. These approaches allow us to not only identify the general performance of each bot detection method but also gain insights into performance variations across each specific dataset. To achieve this, we selectively choose the most accurate models from Table \ref{tab:f1_single_scores}, in addition to the Botometer model \citep{yang2022botometer}. The Botometer model is included because it has already been trained on a diverse range of datasets and is expected to perform well in general-case scenarios.

The results, as presented in Table \ref{tab:f1_all_scores}, indicate that the most generalizable models are BotArtist, Botometer, and SGBot. BotArtist demonstrates superior performance over existing methods in terms of both total and average scores, surpassing almost 10\% the existing methods in total. These results confirm that fine-tuned models which are based on the limited set of features are capable of capturing the general differences between normal and bot accounts.

\section{Conclusions and Future Work}

In this research paper, we introduce a semi-automatic machine learning pipeline (SAMLP) that addresses multiple challenges in creating machine learning models, including feature selection, hyperparameter fine-tuning, model evaluation, decision threshold optimization for binary classification, and provides model explainability through the SHAP game-theoretical approach. We apply this developed pipeline to create BotArtist, a versatile bot detection model based on user profile features. Our approach is trained and evaluated alongside current state-of-the-art solutions using nine different datasets. As demonstrated in our experiments, BotArtist surpasses existing, more complex methods in terms of generalization, achieving almost a 10\% increase in the total F1 score in comparisons of multiple data sets and a 6. 5\% increase in comparisons of separated data sets. Additionally, we offer insights into the final model's decision-making process and the patterns it captures, based on the SHAP game-theoretical approach (available at the repository). 

One of the limitations of the presented methodology is the relatively limited feature set, which might make it susceptible to evasion by future bot accounts that specifically aim to avoid detection based on these features. Further research conducted over a longer period is required to assess the model's effectiveness in detecting bot accounts in the evolving landscape of social networks. 

 For further research, we provide one of the largest labeled datasets containing user profiles correlated with another publicly available dataset of the 127M user posts. Such large labeled data will allow the development of state-of-the-art LLM models to detect bot-generated content in the post-Twitter API era.

\bibliography{aaai25}

\end{document}